\documentclass[onecolumn,showpacs]{revtex4}
\usepackage{amssymb,epsfig,amscd}
\usepackage{graphicx}% Include figure files
\usepackage{dcolumn}% Align table columns on decimal point
\usepackage{bm}% bold math

\usepackage{epsfig}

\newcommand{\ud}{\mathrm{d}}
\newcommand{\dirac}{\partial\llap{$\diagup$\kern-2pt}}

\def\be{\begin{equation}} 
\def\ee{\end{equation}}
\def\bq{\begin{eqnarray}} 
\def\eq{\end{eqnarray}}

\begin{document}

\title{Stability of CFL cores in Hybrid Stars}

\author{G. Pagliara, J. Schaffner-Bielich}

\affiliation{Institut f\"{u}r Theoretische Physik, Goethe Universit\"{a}t,
  D-60438, Frankfurt am Main, Germany}

\begin{abstract} 
We study the equation state of strongly interacting quark matter 
within a NJL-like model in which the chiral condensates and the
color superconducting gaps are computed self-consistently as a
function of the baryon density. A vector interaction term is 
added to the Lagrangian in order to render the quark matter equation of state
stiffer.  For the low density hadronic phase we use a relativistic mean field model. 
The phase transition to quark matter is computed
by a Maxwell construction.  We show that stable CFL cores in hybrid
stars are possible if the superconducting gap is sufficiently large.
Moreover we find stable stellar
configurations in which two phase transitions occur, a first
transition from hadronic matter to 2SC quark matter and a second transition
from 2SC quark matter to CFL quark matter.

\end{abstract}

\maketitle

\section{Introduction}
The possibility that quark matter, and eventually color
superconducting quark matter, is present in the center of neutron stars
has stimulated many theoretical investigations in the last years both
on the modelling of the equation of state (EoS) of quark matter and on
the phenomenological signatures of the presence of quark matter
in neutron stars \cite{Alford:2007xm}. Presently the ``state of the art'' for the EoS of quark matter is
represented by the three-flavor Nambu-Jona-Lasinio (NJL) model in which both the chiral
condensates and the diquark condensates are self-consistently computed
as a function of the chemical potential and temperature. Chemical
equilibrium and charge neutrality (both electric and color charge)
conditions necessary to describe neutron star matter are also imposed in this model
\cite{Steiner:2002gx,Abuki:2004zk,Ruester:2005jc,Blaschke:2005uj,Ippolito:2007uz}. Recently, also the
effect of a finite neutrino chemical potential has been included
\cite{Ruester:2005ib,Sandin:2007zr}. The structure of the QCD phase
diagram within this model turns out to be very rich, with many
different possible quark phases. One of the most striking feature, on which we will focus
here, is the first order phase transition between the two flavor phase,
2SC or normal quark matter depending on the diquark coupling constant,
and the three-flavor superconducting phase, the Color-Flavor-Locking
phase (CFL), at vanishing temperature.
 
Concerning the phenomenological signatures, the differences between the
mass-radius relation for neutron stars and quark or hybrid stars is currently
studied. The astrophysical data so far are 
still affected by large uncertainties but will improve considerably with the advent of new 
satellite missions, as XEUS, Constellation-X, SKA, JWST and LISA. 
From recent theoretical studies it turns out that the values of the maximum
mass of neutron stars and compact stars containing quark matter are very similar
\cite{Alford:2004pf,Alford:2006vz}. Therefore, it seems difficult with
the present knowledge to rule out quark matter from just a mass
measurement \cite{Nice:2005fi,Ozel:2006bv}. Other interesting
quantities have been calculated (see Ref.~\cite{SchaffnerBielich:2007mr}
for a recent review) for the different possible phases of quark matter, as the
neutrino emissivity and the heat capacity which are important for the cooling
of compact stars \cite{Shovkovy:2002kv,Reddy:2002xc,Aguilera:2005tg,Popov:2005xa,Anglani:2006br} 
or the bulk viscosity which determines
stability with respect to gravitational waves emission via r-modes
\cite{Madsen:1999ci,Sa'd:2006qv,Alford:2006gy,Alford:2007rw,Drago:2003wg,Drago:2007iy}. 
Also in explosive phenomena, as supernovae and gamma-ray-bursts (GRBs),
quark matter can play an important role. For instance, the possibility of a
double phase transition, first from hadronic matter to 
2-flavor quark matter and then from 2-flavor quark matter to the CFL phase has been
proposed to explain the complicated time structure of GRBs exhibiting a long
quiescent time in their light curve
\cite{Drago:2005qb,Drago:2005rc,parenti}.

When the above mentioned NJL-EoS is used for the applications on
compact stars, hybrid stars become unstable at the onset of the CFL phase
and therefore CFL phase can not be present in
the core of neutron stars. This conclusion was obtained first in
Ref.~\cite{Baldo:2002ju} where quark matter does not occur at all in compact stars 
because there is a direct transition
from hadronic matter to CFL matter in the model used.  
In
Refs.~\cite{Buballa:2003et,Klahn:2006iw,Grunfeld:2007jt} for different 
EoSs for the hadronic matter and different parameters for the NJL model for
quark matter, again the CFL phase was ruled out because it renders the stars unstable.  
The 2SC phase could appear; the
conclusion is therefore that only 2-flavor superconducting quark matter can be realized
in compact stars. A similar result about the allowed quark phases in compact stars was also found in
Ref.~\cite{Lawley:2006ps} where a modified 2-flavor NJL model, which simulates
confinement at low density, is proposed. 
The dependence of the stability of a quark core 
from the momentum cut-off of the NJL model has been analysed using density
dependent cut-offs, but the instability still persists \cite{Baldo:2006bt} 
\footnote{Very recently, stable hybrid stars configurations have been also obtained
by adopting the three flavor LOFF phase for the quark matter EoS \cite{nicola}.} .

On the other hand, completely different results
are obtained using MIT-bag-like models as shown in
Refs.~\cite{Alford:2002rj,Drago:2004vu,Bombaci:2006cs} where the
appearance of CFL cores does not compromise the stability of the
star. Moreover, the absence of the 2SC phase in compact stars has been
demonstrated in Ref.~\cite{Alford:2002kj}.

In this paper we want to consider again the NJL-EoS as computed in
Ref.~\cite{Ruester:2005jc} to study the structure and composition of
compact stars. We will investigate larger windows of the model
parameters with respect to previous work, with particular attention to
the diquark coupling. Furthermore, we discuss the importance of the
procedure used to fix
the effective bag constant within the NJL model for the stability
of a star when the phase transition to quark matter is considered.  We will
investigate a new procedure for fixing the effective bag constant
by requiring that the chiral symmetry restoration coincides with 
the transition from the hadronic to the quark matter description 
\footnote{In the following, we adopt the term deconfinement transition for the switch 
from the hadronic to the quark model, as usually done in the literature. However, 
the Polyakov loop is not a good order parameter for the deconfinement transition at 
finite chemical potential anymore but only the quark condensate for the chiral phase transition 
\cite{Fraga:2001id,McLerran:2007qj}.}. 
We use then different EoSs to
compute the mass-radius relations of hybrid stars showing that in some
cases a stable CFL core is possible and, even more intriguing, that two phase
transitions, from hadronic matter to the 2SC and then to the CFL phase, can take place in
compact stars. The double phase transition is particular intriguing
in connection with the
deconfinement quark model of GRBs in which the interpretation of bursts
presenting two emission periods is due to a double phase transition in
compact stars \cite{Drago:2005qb,Drago:2005rc,parenti}.

The paper is organised as follows. In Sec.~II we discuss a simple toy
model for a first order phase transition between hadronic matter and quark matter in compact
stars and we study the stability of the star by varying
the parameters of the quark matter EoS. In Sec.~III we compute the quark matter EoS within the
NJL model for different sets of parameters and finally in Sec.~IV we
discuss the stability of the stars obtained using our EoSs with
particular emphasis on the CFL core stability.  In Sec.~V we draw our
conclusions.

\section{A toy-model for phase transitions in compact stars} 

 \begin{figure}
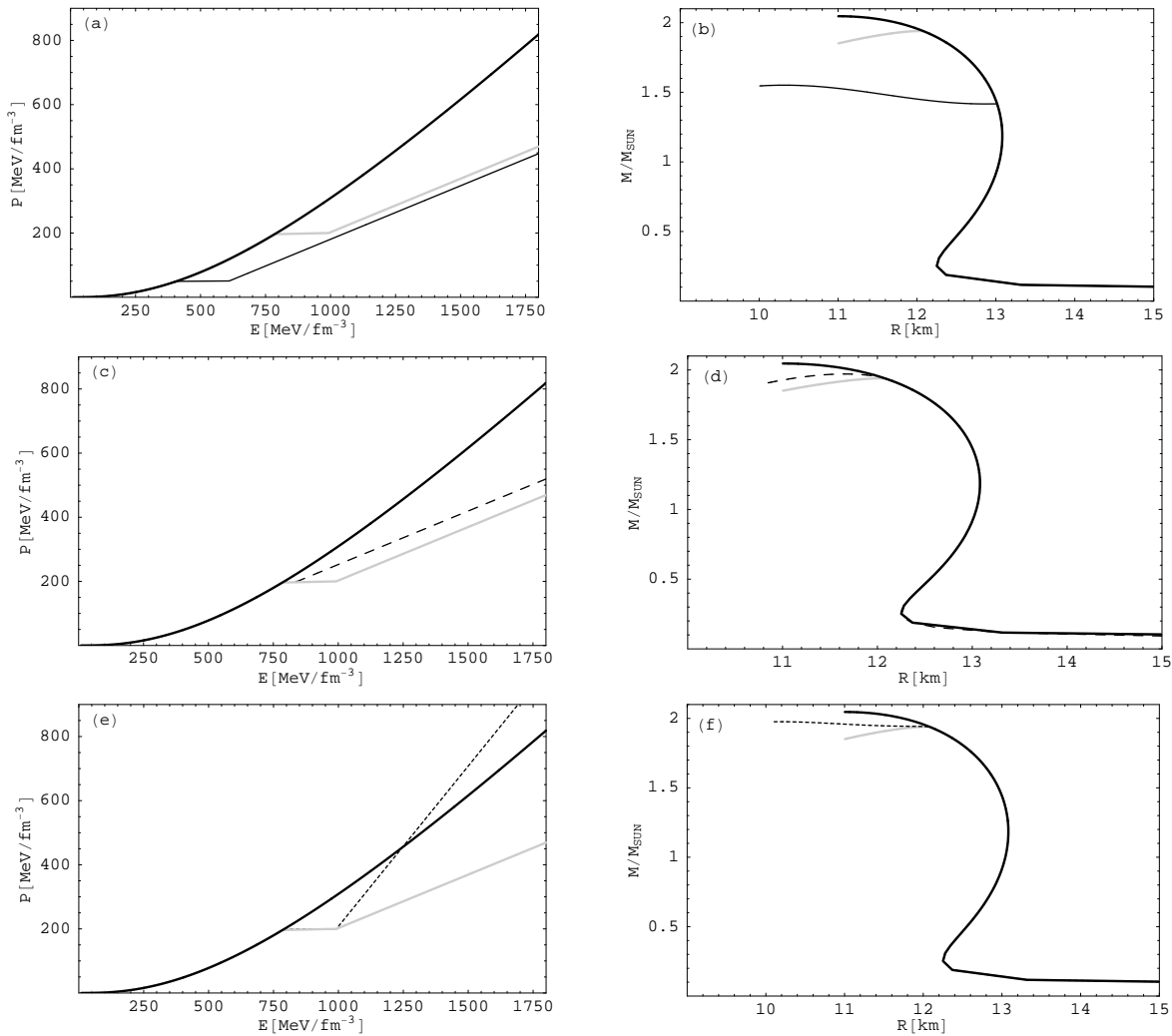

    \begin{centering}
\hbox{\hskip-0.cm \epsfig{file=toy-eos1.epsi,height=4.4cm}\hskip 1.cm 
\epsfig{file=toy-mr1.epsi,height=4.4cm}} \vskip 0.2cm
\hbox{ \hskip-0.cm \epsfig{file=toy-eos2.epsi,height=4.4cm}\hskip 1.cm 
\epsfig{file=toy-mr2.epsi,height=4.4cm}}  \vskip 0.2cm
\hbox{ \hskip-0.cm \epsfig{file=toy-eos3.epsi,height=4.4cm}\hskip 1.cm 
\epsfig{file=toy-mr3.epsi,height=4.4cm}}
    \caption{Equations of state and corresponding mass-radius relations for the toy-model
for different parameter sets. EoSs are shown in the left row and mass-radius relations in the right row.
The purely hadronic matter EoS and the corresponding neutron stars sequence are depicted by a 
thick solid line. Panel (a): quark matter EoS for  
 $a=1/3$, $\Delta_{\epsilon}=200$ MeV/fm$^3$ and
$p_0=200$ MeV/fm$^3$ (gray line set P1), and $a=1/3$, $\Delta_{\epsilon}=200$ MeV/fm$^3$ and
$p_0=50$ MeV/fm$^3$ (solid line set P2). Panel (c): quark matter EoS for the choice
 $a=1/3$, $\Delta_{\epsilon}=50$ MeV/fm$^3$ and
$p_0=200$ MeV/fm$^3$ (dashed line set P3), the other curves are as in panel (a).
Panel (e): quark matter EoS for $a=1$, $\Delta_{\epsilon}=200$ MeV/fm$^3$ and
$p_0=200$ MeV/fm$^3$ (dotted line set P4), the other curves as in panel (a). Stable 
configurations with a quark core can only appear for sufficiently stiff quark matter EoS or 
for a transition point away from the maximum mass configuration of the purely hadronic compact stars.
\label{toy}}
   \end{centering}
   \end{figure}

 \begin{figure}
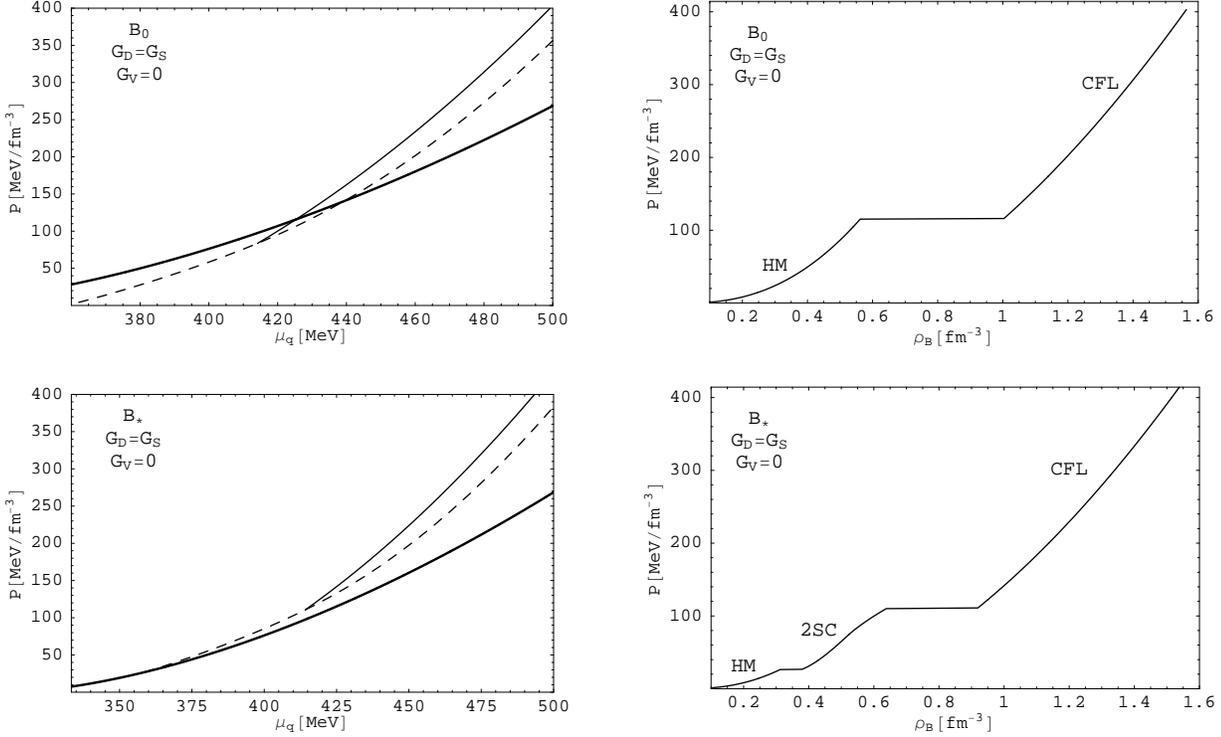

    \begin{centering}
\hbox{\hskip-0.cm \epsfig{file=eos1.epsi,height=4.6cm}\hskip 1.cm 
\epsfig{file=prho1.epsi,height=4.6cm}} \vskip 0.5cm
\hbox{\hskip-0.cm \epsfig{file=eos2.epsi,height=4.6cm}\hskip 1.cm 
\epsfig{file=prho2.epsi,height=4.6cm}}
\caption{Upper left panel: pressure as a function of the chemical potential 
for hadronic matter (thick line), the 2SC phase (dashed line) and the CFL phase (solid line).
Upper right panel: pressure as a function of the baryon density. Parameters are:
$G_D=G_S$, $B=B_0$, $G_V=0$. A direct
transition from hadronic matter to the CFL phase occurs with a large baryon density jump $\sim 0.45$ fm$^{-3}$ 
at the onset of the phase transition.
Lower left panel: pressure as a function of the chemical potential. 
Lower right panel: pressure as a function of the baryon density. Parameters are:
$G_D=G_S$, $B=B_*$, $G_V=0$. For a lower value of $B$ there is first a 
transition from hadronic matter to 2SC quark matter and then from 
2SC quark matter to CFL quark matter.
\label{eos1} }
   \end{centering}
   \end{figure}

We present in this section a toy-model EoS for strongly interacting
matter to show qualitatively the conditions for stable solutions of the
Tolman-Oppenheimer-Volkoff (TOV) equation when a phase
transition from hadronic matter to an exotic phase, like quark matter, occurs.  At very low
baryon densities, $n \lesssim 0.05$fm$^{-3}$, we use the
EoS as computed in Ref.~\cite{Ruester:2005fm} 
(similar results are obtained using the EoS of Ref.~\cite{Baym:1971pw}) 
suitable for the crust of neutron stars. At larger densities, we consider for the hadronic matter a
relativistic mean field EoS, GM3, taken from
Ref.~\cite{Glendenning:1991es}.  For quark matter we adopt a schematic
EoS in which the pressure is proportional to the energy density $p=a
\epsilon $ where the slope (which corresponds to the sound velocity
and therefore regulates the stiffness of the quark matter EoS) is a free
parameter. Notice that this EoS corresponds, for $a=1/3$, to the EoS
of massless and non-interacting quarks. We then model the phase
transition from hadronic matter to quark matter by introducing two other free parameters, the
energy density jump $\Delta_{\epsilon}=\epsilon_2-\epsilon_1$ at the onset of the phase
transition ($\epsilon_2$ and $\epsilon_1$ are the energy densities of the quark phase and of the hadronic phase)  
and the pressure at which the phase transition occurs
$p_0$.  Our aim is to investigate, for the different values of these
parameters, the stability of stars having a core of quark matter.  Let $P1$ be
the parameter set with $a=1/3$, $\Delta_{\epsilon}=200$ MeV/fm$^3$ and
$p_0=200$ MeV/fm$^3$.  In panel (a) of Fig.~\ref{toy} the
corresponding EoS is plotted by a gray line. In this case compact stars
are unstable when the quark matter phase appears, see the gray line in panel (b). If
we decrease now the transition pressure to $p_0=50$ MeV/fm$^3$ keeping
the same values for the other parameters (set $P2$ see solid line in
panel (a)), a sizable branch of stable configurations appears instead, see the
thin solid line in panel (b).  If we reduce the energy density jump to $\Delta_{\epsilon}=50$ MeV/fm$^3$
(set $P3$ with $p_0$ and $a$ as in set $P1$, see dashed line in panel (c)) a small branch of stable
solutions is obtained (dashed line in panel (d)).  Concerning the
stiffness parameter $a$, to obtain stable solutions one must choose $a=1$ (set $P4$ with $p_0$ 
and $\Delta_{\epsilon}$ as in $P1$). For this value of $a$
the quark matter EoS is even stiffer than the hadronic matter EoS (see the dotted line
in panel (e) for the EoS and the dotted line in panel (f) for the
mass-radius relation). This last case is probably not realistic in view
of the asymptotic freedom property of QCD.

In conclusion, what seems to be the most crucial parameter for the
stability of a quark matter core is the value of the pressure at the onset of the phase
transition. If the softening due to the appearance of quark matter occurs at too large pressures,
deep in the core of a neutron star, the sourronding material exerts a
pressure that can not be sustained by the new formed phase and
therefore the star collapses. Also a low value of $\Delta_{\epsilon}$
can help to stabilise the quark matter core. A detailed analytical study
on the critical value of energy density jump for having stable cores of a new phase 
can be found in Refs.~\cite{1971AZh....48..443S,1987A&A...172...95Z,kaempfer,Lindblom:1998dp}
demonstrating that $\epsilon_2/\epsilon_1$ should be not larger than  
$\epsilon_2/\epsilon_1 \leq \frac{3}{2}(1+p_0/\epsilon_1) $.

We will see in the next sections how these cases are connected with more physical meaningful quantities as
the bag constant, the superconducting gap and the constituent masses of quarks .

\section{ Phase transition to quark matter }

We present now a more realistic EoS for quark matter.
The EoS is computed within the NJL-like model proposed
in Ref.~\cite{Ruester:2005jc} in which a scalar diquark interaction term
for the color antitriplet and flavor antitriplet channel is added to the 
usual NJL model. Here we include also the isoscalar vector term 
as in Ref.~\cite{Hanauske:2001nc,Manka:2002yv,Klahn:2006iw} in order to obtain stiffer EoSs.
The input variables of the model are the chemical potentials for all the quark flavors and colors 
given, in chemical equilibrium, by the matrix:
 \begin{equation}
\mu_{ab}^{\alpha\beta} = \left(
  \mu \delta^{\alpha\beta} 
+ \mu_Q Q_{f}^{\alpha\beta} \right)\delta_{ab} 
+ \left[ \mu_3 \left(T_3\right)_{ab} 
+ \mu_8 \left(T_8\right)_{ab} \right] \delta^{\alpha\beta} \; .
\label{mu-f-i}
\end{equation}
where $\mu$ is the quark chemical potential, $\mu_Q$ is the chemical
potential of the electric charge equal to minus the electron
chemical potential $\mu_e$, $\mu_3$ and $\mu_8$ are the color chemical
potentials associated with the two mutually commuting color charges of the
$SU(3)_c$ gauge group. The explicit form of the electric charge matrix
is $Q_{f}=\mbox{diag}_{f}(\frac23,-\frac13,-\frac13)$, and for
the color charge matrices 
$T_3=\mbox{diag}_c(\frac12,-\frac12,0)$ and
$\sqrt{3}T_8=\mbox{diag}_c(\frac12,\frac12,-1)$.

The model parameters are fixed 
by fitting low energy hadronic properties which are the current quark masses,
the quark-antiquark coupling $G_S$, the strength $K$ of  the ``'t Hooft'' interaction and
the cut-off parameter $\Lambda$
introduced in the NJL model to regularise the ultraviolet divergences:

\begin{eqnarray}
m_{u,d} &=& 5.5 \; \mathrm{MeV} \; , \\
m_s &=& 140.7 \; \mathrm{MeV} \; , \\
G_S \Lambda^2 &=& 1.835 \; , \\
K \Lambda^5 &=& 12.36 \; , \\
\Lambda &=& 602.3 \; \mathrm{MeV} \; .
\label{Lambda} 
\end{eqnarray}

After fixing the masses of the up and down quarks by equal values, 
$m_{u,d}=5.5~\mbox{MeV}$, the other four parameters are chosen to 
reproduce the following four observables \cite{Buballa:2003qv}: 
$m_{\pi}=135.0~\mbox{MeV}$, $m_{K}=497.7~\mbox{MeV}$,
$m_{\eta^\prime}=957.8~\mbox{MeV}$, and $f_{\pi}=92.4~\mbox{MeV}$.
This parameter set gives $m_{\eta}=514.8~\mbox{MeV}$ \cite{Buballa:2003qv}.

There are two more parameters, the diquark coupling $G_D$ and
the vector current coupling $G_V$ which are not known. We will use 
$G_D=G_S$ and  $G_D=1.2 G_S$ because
one expects that the diquark coupling has the strength as the 
quark-antiquark coupling.
For $G_V$ we choose the cases $G_V=0$ and  $G_V=0.2G_S$.

At vanishing temperature and within the mean field approximation the pressure reads:

\begin{eqnarray}
p &=& \frac{1}{2 \pi^2} \sum_{i=1}^{18} \int_0^\Lambda \ud k \, k^2
|\epsilon_i|+ 4 K \sigma_u \sigma_d \sigma_s
- \frac{1}{4 G_D} \sum_{c=1}^{3} \left| \Delta_c \right|^2
-2 G_S \sum_{\alpha=1}^{3} \sigma_\alpha^2+ \frac{\omega_0^2}{4 G_V}+ p_e
\label{pressure}
\end{eqnarray}

where $\epsilon_i$ are the dispersion relations as computed in
Ref.~\cite{Ruester:2005jc}, $\sigma_{u,d,s}$ are the quark-antiquark
condensates and $\Delta_c$ are the three diquark condensates. We
denote with $\omega_0=2 G_V\langle QM|\psi_u^{\dagger}
\psi_u+\psi_d^{\dagger} \psi_d +\psi_s^{\dagger} \psi_s |QM\rangle$ the mean field expectation
value of the scalar vector meson $\omega$.
This field modify also the chemical potentials: $\mu_{u,d,s} \rightarrow \mu_{u,d,s}-\omega_0$. 
Finally, the contribution to the pressure of electrons is $p_e=\mu_e^4/(12 \pi^2)$.

The pressure within the NJL model is defined but for a constant $B$, similarly to the MIT bag constant, 
which is usually
fixed by the following procedure: one requires that the corrected pressure $p-B$ is vanishing at
vanishing chemical potential \cite{Buballa:1998pr,Schertler:1999xn,Blaschke:2005uj}. 
In our model, for the parameters set used here, we have
$B=B_0=(425.4 MeV)^4$ \footnote{To compare with
Ref.~\cite{Schertler:1999xn}, consider that within our formalism the
bag constant corresponds to
$B_0=B_0^{ref}+\sum_{u,d,s}\frac{3}{\pi^2}\int_0^\Lambda \ud k \, k^2
\sqrt{k^2+m_i^2} $ where $B_0^{ref}=(217.6 MeV)^4$ from
Ref.~\cite{Schertler:1999xn}. We have checked that our EoS for normal quark matter
is equal to the one presented in that paper. }.  Actually, this
procedure to determine the bag constant is somehow unsatisfying, as
also stated in Ref.~\cite{Schertler:1999xn}, since the
pressure computed within the NJL model at vanishing density is used i.e. in a
regime where NJL model can not be trusted due to its lack of
confinement. On the other hand, in the MIT bag model for instance,
which contains confinement, the pressure in the vacuum is not
vanishing.  

Here we propose an alternative procedure to fix the bag
constant in the NJL model. First, we introduce at low density an EoS
having hadronic degrees of freedom, like the GM3 EoS used in Sec. II,
and then we compute the transition to quark matter, the deconfinement transition,
by a Maxwell construction. We remark that "deconfinement" in our
scheme has the meaning of a change of degrees of freedom and the corresponding Lagrangian,
it is not a phase transition described by an order parameter. 
To fix the bag constant we assume that deconfinement occurs at
the same chemical potential as the chiral phase transition i.e.  when
chiral symmetry is restored. Practically this means that we require
that the pressure of quark matter, $p-B$, is equal to the pressure of the hadronic matter at the critical
chemical potential for which chiral symmetry is restored i.e. the value $\mu_{crit}$ computed
in the NJL model. This allows to fix the value of $B$ and, as we will
see, to obtain significantly different results with respect to the ones
obtained using the conventional procedure. It turns out
that deconfinement occurs at very large quark chemical potential $\sim 470$ MeV
by far larger than the critical chemical potential for chiral
symmetry restoration for $B=B_0$, the standard choice \cite{Schertler:1999xn}. 
The bag value obtained with our assumption,
$B=B_*$, as we will see in the following is marginally smaller than $B_0$ and must be considered as the lowest possible value for the bag
constant in the NJL model because it allows to use the NJL-EoS just starting from $\mu_{crit}$.
For chemical potentials lower than $\mu_{crit}$ the density
of quarks, as computed within the NJL model, is vanishing due to the
completely broken chiral symmetry. This is obviously a regime in which the NJL model
can not be applied.
Our assumption on the coincidence
of deconfinement and the chiral phase transitions at finite density, has not yet a QCD-motivated
argument. Nevertheless, this coincidence has been found in
Lattice QCD calculations at finite temperature (see Ref.~\cite{Hatta:2003ga} and references therein) 
and it has been also adopted in
other models for the EoS at finite chemical potential as the
NJL-inspired model proposed in
Ref.~\cite{Lawley:2006ps}. Interestingly, within the Dyson-Schwinger
approach, it is possible to define an order parameter for
deconfinement at finite density
and it turns out that the two phase transitions occur simultaneously \cite{Bender:1997jf}.
We will examine here both choices for the bag constant $B_0$ and $B_*$.

We remark that another possible scenario has been proposed for the finite density phase
transition in Refs.~\cite{Fraga:2001id,McLerran:2007qj}: 
there is no deconfinement at all at large density, only
the chiral phase transition occurs and the quarks are still confined.

\subsection{Results}

In order to compute the EoS needed for compact stars,
the pressure, Eq.~(\ref{pressure}), must be minimised with respect
to the chiral and color superconducting order parameters,
$\sigma_\alpha$ and $\Delta_c$, and therefore six gap equations are obtained:

\begin{eqnarray}
\frac{\partial p}{\partial \sigma_\alpha} &=& 0 \;  , \\
\frac{\partial p}{\partial \Delta_c} &=& 0 \; .
\end{eqnarray}

Moreover, local electric and color charge neutrality are met
if three other three equations are satisfied,

\begin{eqnarray}
n_Q &\equiv& \frac{ \partial p }{\partial \mu_Q} = 0 \; , \\
n_3 &\equiv& \frac{ \partial p }{\partial \mu_3} = 0 \; , \\
n_8 &\equiv& \frac{ \partial p }{\partial \mu_8} = 0 \; .
\end{eqnarray}

These conditions fix the values of the three corresponding chemical potentials,
$\mu_Q$, $\mu_3$ and $\mu_8$. One more equation is imposed
to compute the vector current expectation value:

\begin{equation}
\frac{\partial p}{\partial \omega_0} = 0. 
\end{equation}

 \begin{figure}
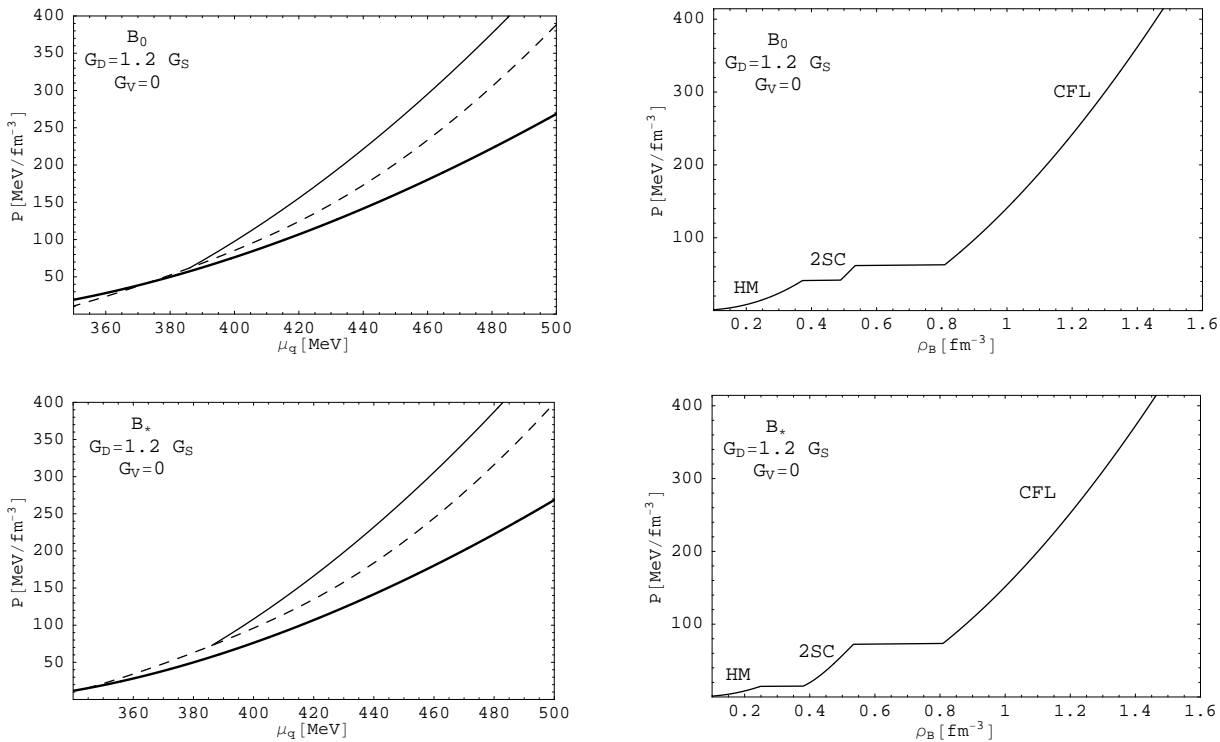

    \begin{centering}
\hbox{\hskip-0.cm \epsfig{file=eos3.epsi,height=4.6cm}\hskip 1.cm 
\epsfig{file=prho3.epsi,height=4.6cm}} \vskip 0.5cm
\hbox{\hskip-0.cm \epsfig{file=eos4.epsi,height=4.6cm}\hskip 1.cm 
\epsfig{file=prho4.epsi,height=4.6cm}}
    \caption{Same as in Fig.~\ref{eos1} for 
$G_D=1.2 G_S$. When a large diquark coupling $G_D$ is
considered there are for both values of $B$ two phase transitions,
first from hadronic matter to 2SC quark matter
and then from 2SC quark matter to CFL quark matter.
\label{eos3} }
   \end{centering}
   \end{figure}

After all these quantities are fixed, one arrives at the pressure of quark matter as a
function of the quark chemical potential only and one can easily build
the Maxwell construction by solving the equation:
\begin{equation} 
p_{HM}(\mu)=p_{QM}(\mu)-B. 
\end{equation} 
As discussed before we use for $B$ two values, $B_0$
and $B_*$.  In the upper left panel of Fig.~\ref{eos1} we show the
pressure as function of the quark chemical potential for hadronic matter 
(thick line) and  quark matter (thin line, the dashed line corresponding
to the 2SC phase and the continuous line to the CFL phase). Parameters are
$B=B_0$, $G_D=G_S$ and $G_V=0$. In this case there is a direct
transition from hadronic matter to the CFL phase, see also Refs.~\cite{Baldo:2002ju,Buballa:2003et}. 
The corresponding EoS is
shown in the right panel of Fig.~\ref{eos1}. Notice the large jump of
the baryon density of about $\sim 0.45 $ fm$^{-3}$ at the onset of the phase
transition. We notice that in Ref.~\cite{Klahn:2006iw} a similar 
parameter set, $B=B_0$, $G_D=GS$ and $G_V=0$, gives a different result:
there is just a transition from hadronic matter to the 2SC phase.  Apart from the
different choice for the hadronic matter EoS, the 't Hooft interaction term is
neglected in that calculation. We obtain the same result, a transition from hadronic matter to the 2SC phase, by 
choosing $K=0$ in our
model. The 't Hooft term in fact, as observed in
Refs.\cite{Buballa:2003qv,Drago:2007zk}, pushes  the
2SC-CFL phase transition to lower chemical potentials rendering the CFL phase the favored quark matter phase also
at intermediate densities.  

Let us study how this result changes
if we choose $B=B_*$, which for this set of parameters is $B_*=(424.8
MeV)^4$, slightly smaller than $B_0=(425.4 MeV)^4$. 
By construction now there is first a transition from the hadronic matter
to the 2SC matter and then a second transition from 2SC matter to CFL
matter, see the corresponding plots in the lower panel of Fig.~\ref{eos1}.

We repeated the previous calculation for $G_D=1.2G_S$. For this value of
the diquark coupling the superconducting gap within the CFL phase at
$\mu=500$ MeV is $\Delta_{CFL}\sim 160$ MeV. Notice that our present
knowledge of the CFL gap concerns only its order of magnitude
i.e. $\sim 100$ MeV; therefore, a CFL gap larger than $100$ MeV is not excluded and
has been considered also in previous papers
\cite{Neumann:2002jm,Alford:2002rj,Blaschke:2005uj,Kitazawa:2007zs}. The effect of
increasing the diquark coupling on the EoS is to decrease the onset
of chiral symmetry restoration, from $\mu_q=358$ MeV for $G_D=G_S$ to $\mu_q=344 $
MeV for $G_D=1.2G_S$ and the 2SC-CFL phase transition onset, from $\mu_q=415$
MeV for $G_D=G_S$ to $\mu_q=386 $ MeV for $G_D=1.2G_S$. As shown in
Fig.~\ref{eos3} for both choices of $B$ there is a
double phase transition with increasing baryon density \footnote{It
is interesting to notice that in the cases in which a transition from
hadronic matter to the 2SC phase is found, the two order parameters represented by the
chiral condensate and the superconducting gap both abruptly change at the same chemical potential.
It is therefore interesting to investigate, in a model independent way, whether the superconducting gap can be considered as the order parameter
of confinement at finite density.}.

\begin{figure}
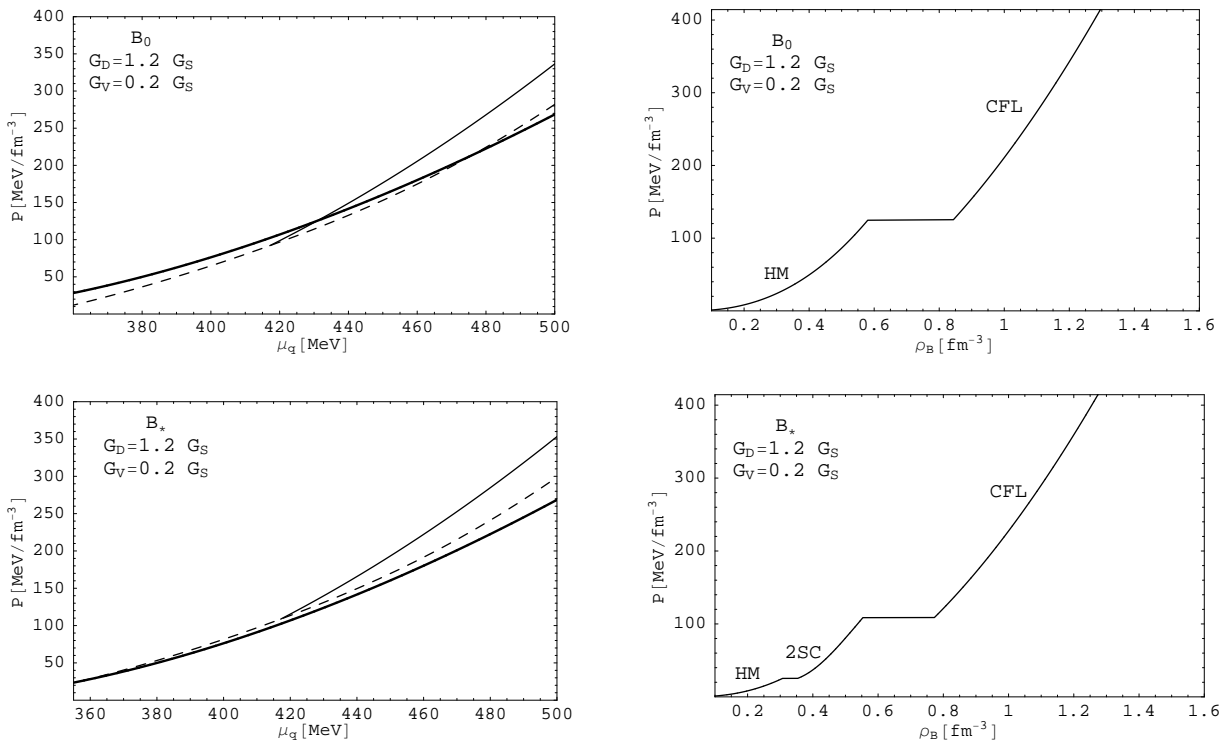

    \begin{centering}
\hbox{\hskip-0.cm \epsfig{file=eos-vec1.epsi,height=4.6cm}\hskip 1.cm 
\epsfig{file=prho-vec1.epsi,height=4.6cm}} \vskip 0.5cm
\hbox{\hskip-0.cm \epsfig{file=eos-vec2.epsi,height=4.6cm}\hskip 1.cm 
\epsfig{file=prho-vec2.epsi,height=4.6cm}}
\caption{Same as in Fig.~\ref{eos3} with the inclusion of a repulsive vector term with 
$G_V=0.2 G_S$. In the case $B=B_0$ a direct
transition from hadronic matter to CFL matter is found. In the case $B=B_*$ the EoS
exhibits instead two phase transitions. 
\label{eos5} }
   \end{centering}
   \end{figure}

We include now in our calculation the vector meson term and set
$G_V=0.2G_S$ (we consider now only $G_D=1.2G_S$). The physical effect of this term is
a repulsive interaction between quarks which renders the quark matter EoS stiffer. 
For the case $B=B_0$, there is
a transition from hadronic matter to CFL matter and for the case $B=B_*$ a
double phase transition is present, see Fig.~\ref{eos5}.

\section{Mass-radius relations}
\begin{figure}
    \begin{centering}
\epsfig{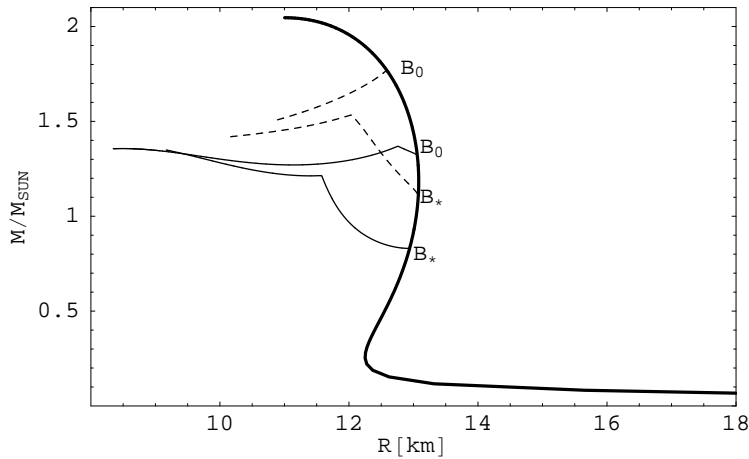}
    \caption{Mass-radius relations for the EoSs with $G_D=G_S$ (dashed lines) for $B=B_0$ and $B=B_*$
and with $G_D=1.2 G_S$ (solid lines) for $B=B_0$ and $B=B_*$. $G_V=0$ in all cases. The thick curve stands
for the hadronic stars. In the case $G_D=G_S$ no stable CFL cores are
found, only a 2SC core is possible for $B=B_*$. In the case $G_D=1.2
G_S$, hybrid stars having a CFL core are stable. Furthermore, in both cases, $B=B_0$
and $B=B_*$ a layer of 2SC phase is present inside the hybrid stars.
\label{mr1} }
   \end{centering}
   \end{figure}

Let us now discuss the corresponding mass-radius relations for compact stars.  In
Fig.~\ref{mr1} we show the mass-radius diagram for the EoSs with
$G_D=G_S$ for $B=B_0$ (EoS1) and $B=B_*$ (EoS2) indicated by the dotted
lines and the EoSs with $G_D=1.2G_S$ for $B=B_0$ (EoS3) and $B=B_*$
(EoS4) indicated by the solid lines.  In the first case, EoS1, where a
transition from hadronic matter to CFL is found (see upper panel of Fig.~\ref{eos1}), the CFL core is
unstable and therefore the conclusion is that quark matter does not occur at all
in compact stars in agreement with the findings of
Ref.~\cite{Buballa:2003et}. This is due to the fact that the transition
to CFL matter occurs at a large pressure and chemical potential which,
as demonstrated within the toy model of Sec.~I, strongly disfavours stable
configurations. Decreasing the value of $B$ to $B_*$, which in the toy
model would correspond to a change of the transition pressure,
changes the result significantly: a stable core of quark matter can be present but only in the 2SC
phase, the subsequent transition to the CFL phase renders the star
unstable. This scenario agrees with the one proposed for instance
in Ref.\cite{Klahn:2006iw}.  Let us discuss now the cases in which
$G_D=1.2G_S$. Obviously the larger the diquark coupling the more
favoured is quark matter with respect to hadronic matter because both the onset of the chiral symmetry 
restoration and the one for the 2SC-CFL transition are shifted to lower densities. 
This implies that the pressure of
the phase transition is smaller than in the previous cases but also that
the jump in density due to the Maxwell construction is
larger \footnote{Notice that increasing the diquark coupling
increases the slope of the pressure as a function of the chemical
potential and therefore also the baryon density.}
. Among these two effects, the first favors the stability of the
star while the second disfavors the stability of the star (see our
toy-model discussion). We find that the first effect dominates
and hybrid stars containing both
the 2SC phase and the CFL phase are stable in both cases, $B=B_0$ and $B=B_*$ 
\footnote{We point out that in the case $B=B_0$, 
the branch of stars containing the CFL phase exhibits first a region of unstable 
solutions and then stable solutions. 
These new stable solutions represent stars belonging to the so-called 
"third family" of compact stars found in Ref.~\cite{Glendenning:1998ag,Schertler:2000xq}.}. 
The structure of these
stellar objects is extremely interesting: it contains a crust of hadronic matter matter,
a layer of 2SC phase and a core of CFL phase. Possible astrophysical
implications of these stellar compositions have been already discussed
in Refs.~\cite{Drago:2005qb,Drago:2005rc,parenti} in connection with GRBs. The
formation of first the 2SC phase
and then of the CFL phase, during the evolution of the star,
would produce two separate neutrino
emission and possibly also two different GRBs emission
periods for which there are already observational hints
\cite{Drago:2005rc}.

\begin{figure}
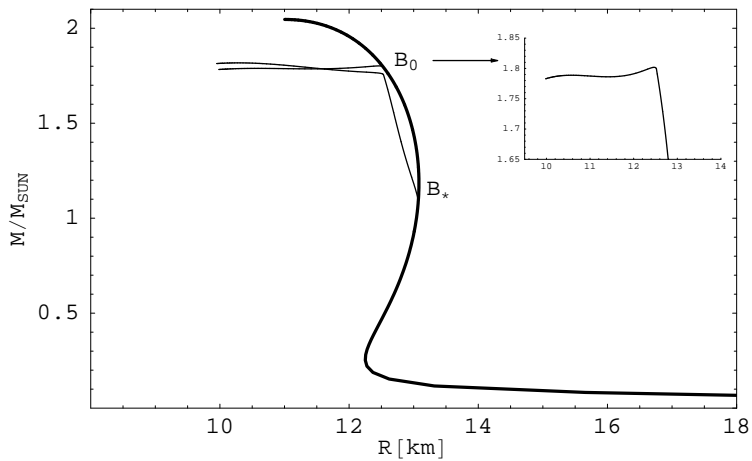

    \begin{centering}
\hbox{\hskip 4cm \epsfig{file=mr-vec.epsi,height=6cm}\hskip 1.cm} \vskip -5.7cm
\hbox{\hskip 10.5cm \epsfig{file=insert.epsi,height=1.8cm}\hskip 1.cm }\vskip 3.5cm
\caption{Mass-radius relations for the EoSs with $G_D=1.2 G_S$ and $G_V=0.2 G_S$ 
for $B=B_0$ and $B=B_*$ (solid lines). The thick curve stands for the
hadronic stars. In both cases stars having a CFL core are stable (see the magnification for the case $B=B_0$), in
the case $B=B_*$ a double phase transition is present.
\label{mr2} }
   \end{centering}
   \end{figure}

We point out that the solutions of the TOV equations obtained 
without vector interactions reach a 
maximum mass of less than $1.4 M_\odot$
and therefore are excluded by pulsar mass data.
For the case $G_V=0.2G_S$ we obtain hybrid stars with a
stable core of CFL phase in the case $B=B_0$ and hybrid stars 
with both the
2SC and the CFL phases for $B=B_*$ as in the previous cases.
These solutions have a corresponding
maximum mass of $\sim 1.8 M_\odot$ (see Fig.~\ref{mr2}) due
to the effect of the repulsion given by the vector interaction and 
are not ruled out
by the presently available astrophysical data.
For even larger values of $G_V$, we arrive at stable hybrid stars with just a CFL core.

A final important remark concerns the large value of the diquark coupling that
we must use to obtain stable quark matter cores. The crucial quantity for the stability is a low value
of the pressure of the phase transition to quark matter as seen already on our discussion
of toy-model. A low transition pressure can be obtained
for a large diquark coupling but also for a reduced
constituent mass of the
strange quark. Within the NJL model discussed here, $m_s$ turns out to be
quite large. Other quark matter studies within the Dyson-Schwinger approach point
towards lower values of $m_s$  and 
consequently the CFL phase dominates in an enlarged density region
\cite{Nickel:2006kc}. Unfortunately, the EoS of quark matter within the
Dyson-Schwinger approach is not yet available. Our analysis suggests that the corresponding
results for the TOV solutions can be very similar to the ones 
obtained here.

\section{Conclusions}

We have studied the phase transition from hadronic matter to quark matter and the corresponding
impact on the mass-radius relation of compact stars. By using first a toy-model for the quark matter
equation of state we have analysed how the stability of hybrid stars depends on the 
properties of the phase transition. In particular a low value of the pressure at the onset of the phase transition 
seems to be the most crucial quantity for the stability of the hybrid star.
We have then computed the equation of state of quark matter within the NJL model by including 
effects from the chiral condensates, the diquark coupling pattern and a repulsion vector term.
For large enough values of the diquark coupling strength, hybrid stars containing
a CFL core are found to be stable. Moreover, exciting stable hybrid stars containing both a layer of a 2SC phase and a core 
of a CFL phase appears to be possible.
This result opens the possibility to find new signatures of the presence of quark matter in 
compact stars. The formation of different quark phases in compact stars can release a huge amount of energy powering
both energetic bursts of neutrino and gamma-ray signals. A detailed study of the time evolution of a proto-neutron star
including the possible appearance of the 2SC phase followed by the formation of the CFL phase could reveal some
new features and opportunities for detecting the chiral phase transition of QCD in the sky.

\vspace{0.15cm}
{\bf Acknowledgements}:\\
We thank S. R\"{u}ster and D. Rischke for providing us the code to compute the equation of state of quark matter.
G.P. thanks M. Buballa and D. Nickel  for useful discussions and 
the INFN for financial support.
\vspace{0.5cm}

\bibliography{references}

\begin{thebibliography}{60}
\expandafter\ifx\csname natexlab\endcsname\relax\def\natexlab#1{#1}\fi
\expandafter\ifx\csname bibnamefont\endcsname\relax
  \def\bibnamefont#1{#1}\fi
\expandafter\ifx\csname bibfnamefont\endcsname\relax
  \def\bibfnamefont#1{#1}\fi
\expandafter\ifx\csname citenamefont\endcsname\relax
  \def\citenamefont#1{#1}\fi
\expandafter\ifx\csname url\endcsname\relax
  \def\url#1{\texttt{#1}}\fi
\expandafter\ifx\csname urlprefix\endcsname\relax\def\urlprefix{URL }\fi
\providecommand{\bibinfo}[2]{#2}
\providecommand{\eprint}[2][]{\url{#2}}

\bibitem[{\citenamefont{Alford et~al.}(2007{\natexlab{a}})\citenamefont{Alford,
  Rajagopal, Schaefer, and Schmitt}}]{Alford:2007xm}
\bibinfo{author}{\bibfnamefont{M.~G.} \bibnamefont{Alford}},
  \bibinfo{author}{\bibfnamefont{K.}~\bibnamefont{Rajagopal}},
  \bibinfo{author}{\bibfnamefont{T.}~\bibnamefont{Schaefer}}, \bibnamefont{and}
  \bibinfo{author}{\bibfnamefont{A.}~\bibnamefont{Schmitt}}
  (\bibinfo{year}{2007}{\natexlab{a}}), \eprint{arXiv:0709.4635 [hep-ph]}.

\bibitem[{\citenamefont{Abuki et~al.}(2005)\citenamefont{Abuki, Kitazawa, and
  Kunihiro}}]{Abuki:2004zk}
\bibinfo{author}{\bibfnamefont{H.}~\bibnamefont{Abuki}},
  \bibinfo{author}{\bibfnamefont{M.}~\bibnamefont{Kitazawa}}, \bibnamefont{and}
  \bibinfo{author}{\bibfnamefont{T.}~\bibnamefont{Kunihiro}},
  \bibinfo{journal}{Phys. Lett.} \textbf{\bibinfo{volume}{B615}},
  \bibinfo{pages}{102} (\bibinfo{year}{2005}), \eprint{hep-ph/0412382}.

\bibitem[{\citenamefont{Ruester et~al.}(2005)\citenamefont{Ruester, Werth,
  Buballa, Shovkovy, and Rischke}}]{Ruester:2005jc}
\bibinfo{author}{\bibfnamefont{S.~B.} \bibnamefont{Ruester}},
  \bibinfo{author}{\bibfnamefont{V.}~\bibnamefont{Werth}},
  \bibinfo{author}{\bibfnamefont{M.}~\bibnamefont{Buballa}},
  \bibinfo{author}{\bibfnamefont{I.~A.} \bibnamefont{Shovkovy}},
  \bibnamefont{and} \bibinfo{author}{\bibfnamefont{D.~H.}
  \bibnamefont{Rischke}}, \bibinfo{journal}{Phys. Rev.}
  \textbf{\bibinfo{volume}{D72}}, \bibinfo{pages}{034004}
  (\bibinfo{year}{2005}), \eprint{hep-ph/0503184}.

\bibitem[{\citenamefont{Blaschke et~al.}(2005)\citenamefont{Blaschke,
  Fredriksson, Grigorian, Oztas, and Sandin}}]{Blaschke:2005uj}
\bibinfo{author}{\bibfnamefont{D.}~\bibnamefont{Blaschke}},
  \bibinfo{author}{\bibfnamefont{S.}~\bibnamefont{Fredriksson}},
  \bibinfo{author}{\bibfnamefont{H.}~\bibnamefont{Grigorian}},
  \bibinfo{author}{\bibfnamefont{A.~M.} \bibnamefont{Oztas}}, \bibnamefont{and}
  \bibinfo{author}{\bibfnamefont{F.}~\bibnamefont{Sandin}},
  \bibinfo{journal}{Phys. Rev.} \textbf{\bibinfo{volume}{D72}},
  \bibinfo{pages}{065020} (\bibinfo{year}{2005}), \eprint{hep-ph/0503194}.

\bibitem[{\citenamefont{Ippolito
  et~al.}(2007{\natexlab{a}})\citenamefont{Ippolito, Nardulli, and
  Ruggieri}}]{Ippolito:2007uz}
\bibinfo{author}{\bibfnamefont{N.~D.} \bibnamefont{Ippolito}},
  \bibinfo{author}{\bibfnamefont{G.}~\bibnamefont{Nardulli}}, \bibnamefont{and}
  \bibinfo{author}{\bibfnamefont{M.}~\bibnamefont{Ruggieri}},
  \bibinfo{journal}{JHEP} \textbf{\bibinfo{volume}{04}}, \bibinfo{pages}{036}
  (\bibinfo{year}{2007}{\natexlab{a}}), \eprint{hep-ph/0701113}.

\bibitem[{\citenamefont{Steiner et~al.}(2002)\citenamefont{Steiner, Reddy, and
  Prakash}}]{Steiner:2002gx}
\bibinfo{author}{\bibfnamefont{A.~W.} \bibnamefont{Steiner}},
  \bibinfo{author}{\bibfnamefont{S.}~\bibnamefont{Reddy}}, \bibnamefont{and}
  \bibinfo{author}{\bibfnamefont{M.}~\bibnamefont{Prakash}},
  \bibinfo{journal}{Phys. Rev.} \textbf{\bibinfo{volume}{D66}},
  \bibinfo{pages}{094007} (\bibinfo{year}{2002}), \eprint{hep-ph/0205201}.

\bibitem[{\citenamefont{Ruester
  et~al.}(2006{\natexlab{a}})\citenamefont{Ruester, Werth, Buballa, Shovkovy,
  and Rischke}}]{Ruester:2005ib}
\bibinfo{author}{\bibfnamefont{S.~B.} \bibnamefont{Ruester}},
  \bibinfo{author}{\bibfnamefont{V.}~\bibnamefont{Werth}},
  \bibinfo{author}{\bibfnamefont{M.}~\bibnamefont{Buballa}},
  \bibinfo{author}{\bibfnamefont{I.~A.} \bibnamefont{Shovkovy}},
  \bibnamefont{and} \bibinfo{author}{\bibfnamefont{D.~H.}
  \bibnamefont{Rischke}}, \bibinfo{journal}{Phys. Rev.}
  \textbf{\bibinfo{volume}{D73}}, \bibinfo{pages}{034025}
  (\bibinfo{year}{2006}{\natexlab{a}}), \eprint{hep-ph/0509073}.

\bibitem[{\citenamefont{Sandin and Blaschke}(2007)}]{Sandin:2007zr}
\bibinfo{author}{\bibfnamefont{F.}~\bibnamefont{Sandin}} \bibnamefont{and}
  \bibinfo{author}{\bibfnamefont{D.}~\bibnamefont{Blaschke}},
  \bibinfo{journal}{Phys. Rev.} \textbf{\bibinfo{volume}{D75}},
  \bibinfo{pages}{125013} (\bibinfo{year}{2007}), \eprint{astro-ph/0701772}.

\bibitem[{\citenamefont{Alford et~al.}(2005)\citenamefont{Alford, Braby, Paris,
  and Reddy}}]{Alford:2004pf}
\bibinfo{author}{\bibfnamefont{M.}~\bibnamefont{Alford}},
  \bibinfo{author}{\bibfnamefont{M.}~\bibnamefont{Braby}},
  \bibinfo{author}{\bibfnamefont{M.~W.} \bibnamefont{Paris}}, \bibnamefont{and}
  \bibinfo{author}{\bibfnamefont{S.}~\bibnamefont{Reddy}},
  \bibinfo{journal}{Astrophys. J.} \textbf{\bibinfo{volume}{629}},
  \bibinfo{pages}{969} (\bibinfo{year}{2005}), \eprint{nucl-th/0411016}.

\bibitem[{\citenamefont{Alford et~al.}(2007{\natexlab{b}})}]{Alford:2006vz}
\bibinfo{author}{\bibfnamefont{M.}~\bibnamefont{Alford}} \bibnamefont{et~al.},
  \bibinfo{journal}{Nature} \textbf{\bibinfo{volume}{445}}, \bibinfo{pages}{E7}
  (\bibinfo{year}{2007}{\natexlab{b}}), \eprint{astro-ph/0606524}.

\bibitem[{\citenamefont{Ozel}(2006)}]{Ozel:2006bv}
\bibinfo{author}{\bibfnamefont{F.}~\bibnamefont{Ozel}},
  \bibinfo{journal}{Nature} \textbf{\bibinfo{volume}{441}},
  \bibinfo{pages}{1115} (\bibinfo{year}{2006}).

\bibitem[{\citenamefont{Nice et~al.}(2005)}]{Nice:2005fi}
\bibinfo{author}{\bibfnamefont{D.~J.} \bibnamefont{Nice}} \bibnamefont{et~al.},
  \bibinfo{journal}{Astrophys. J.} \textbf{\bibinfo{volume}{634}},
  \bibinfo{pages}{1242} (\bibinfo{year}{2005}), \eprint{astro-ph/0508050}.

\bibitem[{\citenamefont{Schaffner-Bielich}(2007)}]{SchaffnerBielich:2007mr}
\bibinfo{author}{\bibfnamefont{J.}~\bibnamefont{Schaffner-Bielich}}
  (\bibinfo{year}{2007}), \eprint{arXiv:0709.1043 [astro-ph]}.

\bibitem[{\citenamefont{Popov et~al.}(2006)\citenamefont{Popov, Grigorian, and
  Blaschke}}]{Popov:2005xa}
\bibinfo{author}{\bibfnamefont{S.}~\bibnamefont{Popov}},
  \bibinfo{author}{\bibfnamefont{H.}~\bibnamefont{Grigorian}},
  \bibnamefont{and} \bibinfo{author}{\bibfnamefont{D.}~\bibnamefont{Blaschke}},
  \bibinfo{journal}{Phys. Rev.} \textbf{\bibinfo{volume}{C74}},
  \bibinfo{pages}{025803} (\bibinfo{year}{2006}), \eprint{nucl-th/0512098}.

\bibitem[{\citenamefont{Shovkovy and Ellis}(2002)}]{Shovkovy:2002kv}
\bibinfo{author}{\bibfnamefont{I.~A.} \bibnamefont{Shovkovy}} \bibnamefont{and}
  \bibinfo{author}{\bibfnamefont{P.~J.} \bibnamefont{Ellis}},
  \bibinfo{journal}{Phys. Rev.} \textbf{\bibinfo{volume}{C66}},
  \bibinfo{pages}{015802} (\bibinfo{year}{2002}), \eprint{hep-ph/0204132}.

\bibitem[{\citenamefont{Reddy et~al.}(2003)\citenamefont{Reddy, Sadzikowski,
  and Tachibana}}]{Reddy:2002xc}
\bibinfo{author}{\bibfnamefont{S.}~\bibnamefont{Reddy}},
  \bibinfo{author}{\bibfnamefont{M.}~\bibnamefont{Sadzikowski}},
  \bibnamefont{and}
  \bibinfo{author}{\bibfnamefont{M.}~\bibnamefont{Tachibana}},
  \bibinfo{journal}{Nucl. Phys.} \textbf{\bibinfo{volume}{A714}},
  \bibinfo{pages}{337} (\bibinfo{year}{2003}), \eprint{nucl-th/0203011}.

\bibitem[{\citenamefont{Aguilera et~al.}(2005)\citenamefont{Aguilera, Blaschke,
  Buballa, and Yudichev}}]{Aguilera:2005tg}
\bibinfo{author}{\bibfnamefont{D.~N.} \bibnamefont{Aguilera}},
  \bibinfo{author}{\bibfnamefont{D.}~\bibnamefont{Blaschke}},
  \bibinfo{author}{\bibfnamefont{M.}~\bibnamefont{Buballa}}, \bibnamefont{and}
  \bibinfo{author}{\bibfnamefont{V.~L.} \bibnamefont{Yudichev}},
  \bibinfo{journal}{Phys. Rev.} \textbf{\bibinfo{volume}{D72}},
  \bibinfo{pages}{034008} (\bibinfo{year}{2005}), \eprint{hep-ph/0503288}.

\bibitem[{\citenamefont{Anglani et~al.}(2006)\citenamefont{Anglani, Nardulli,
  Ruggieri, and Mannarelli}}]{Anglani:2006br}
\bibinfo{author}{\bibfnamefont{R.}~\bibnamefont{Anglani}},
  \bibinfo{author}{\bibfnamefont{G.}~\bibnamefont{Nardulli}},
  \bibinfo{author}{\bibfnamefont{M.}~\bibnamefont{Ruggieri}}, \bibnamefont{and}
  \bibinfo{author}{\bibfnamefont{M.}~\bibnamefont{Mannarelli}},
  \bibinfo{journal}{Phys. Rev.} \textbf{\bibinfo{volume}{D74}},
  \bibinfo{pages}{074005} (\bibinfo{year}{2006}), \eprint{hep-ph/0607341}.

\bibitem[{\citenamefont{Drago et~al.}(2007{\natexlab{a}})\citenamefont{Drago,
  Pagliara, and Parenti}}]{Drago:2007iy}
\bibinfo{author}{\bibfnamefont{A.}~\bibnamefont{Drago}},
  \bibinfo{author}{\bibfnamefont{G.}~\bibnamefont{Pagliara}}, \bibnamefont{and}
  \bibinfo{author}{\bibfnamefont{I.}~\bibnamefont{Parenti}}
  (\bibinfo{year}{2007}{\natexlab{a}}), \eprint{arXiv:0704.1510 [astro-ph]}.

\bibitem[{\citenamefont{Madsen}(2000)}]{Madsen:1999ci}
\bibinfo{author}{\bibfnamefont{J.}~\bibnamefont{Madsen}},
  \bibinfo{journal}{Phys. Rev. Lett.} \textbf{\bibinfo{volume}{85}},
  \bibinfo{pages}{10} (\bibinfo{year}{2000}), \eprint{astro-ph/9912418}.

\bibitem[{\citenamefont{Sa'd et~al.}(2007)\citenamefont{Sa'd, Shovkovy, and
  Rischke}}]{Sa'd:2006qv}
\bibinfo{author}{\bibfnamefont{B.~A.} \bibnamefont{Sa'd}},
  \bibinfo{author}{\bibfnamefont{I.~A.} \bibnamefont{Shovkovy}},
  \bibnamefont{and} \bibinfo{author}{\bibfnamefont{D.~H.}
  \bibnamefont{Rischke}}, \bibinfo{journal}{Phys. Rev.}
  \textbf{\bibinfo{volume}{D75}}, \bibinfo{pages}{065016}
  (\bibinfo{year}{2007}), \eprint{astro-ph/0607643}.

\bibitem[{\citenamefont{Alford and Schmitt}(2007)}]{Alford:2006gy}
\bibinfo{author}{\bibfnamefont{M.~G.} \bibnamefont{Alford}} \bibnamefont{and}
  \bibinfo{author}{\bibfnamefont{A.}~\bibnamefont{Schmitt}},
  \bibinfo{journal}{J. Phys.} \textbf{\bibinfo{volume}{G34}},
  \bibinfo{pages}{67} (\bibinfo{year}{2007}), \eprint{nucl-th/0608019}.

\bibitem[{\citenamefont{Alford et~al.}(2007{\natexlab{c}})\citenamefont{Alford,
  Braby, Reddy, and Schafer}}]{Alford:2007rw}
\bibinfo{author}{\bibfnamefont{M.~G.} \bibnamefont{Alford}},
  \bibinfo{author}{\bibfnamefont{M.}~\bibnamefont{Braby}},
  \bibinfo{author}{\bibfnamefont{S.}~\bibnamefont{Reddy}}, \bibnamefont{and}
  \bibinfo{author}{\bibfnamefont{T.}~\bibnamefont{Schafer}},
  \bibinfo{journal}{Phys. Rev.} \textbf{\bibinfo{volume}{C75}},
  \bibinfo{pages}{055209} (\bibinfo{year}{2007}{\natexlab{c}}),
  \eprint{nucl-th/0701067}.

\bibitem[{\citenamefont{Drago et~al.}(2005)\citenamefont{Drago, Lavagno, and
  Pagliara}}]{Drago:2003wg}
\bibinfo{author}{\bibfnamefont{A.}~\bibnamefont{Drago}},
  \bibinfo{author}{\bibfnamefont{A.}~\bibnamefont{Lavagno}}, \bibnamefont{and}
  \bibinfo{author}{\bibfnamefont{G.}~\bibnamefont{Pagliara}},
  \bibinfo{journal}{Phys. Rev.} \textbf{\bibinfo{volume}{D71}},
  \bibinfo{pages}{103004} (\bibinfo{year}{2005}), \eprint{astro-ph/0312009}.

\bibitem[{\citenamefont{Drago et~al.}(2006)\citenamefont{Drago, Lavagno, and
  Pagliara}}]{Drago:2005qb}
\bibinfo{author}{\bibfnamefont{A.}~\bibnamefont{Drago}},
  \bibinfo{author}{\bibfnamefont{A.}~\bibnamefont{Lavagno}}, \bibnamefont{and}
  \bibinfo{author}{\bibfnamefont{G.}~\bibnamefont{Pagliara}},
  \bibinfo{journal}{Nucl. Phys.} \textbf{\bibinfo{volume}{A774}},
  \bibinfo{pages}{823} (\bibinfo{year}{2006}), \eprint{astro-ph/0510018}.

\bibitem[{\citenamefont{Drago and Pagliara}(2007)}]{Drago:2005rc}
\bibinfo{author}{\bibfnamefont{A.}~\bibnamefont{Drago}} \bibnamefont{and}
  \bibinfo{author}{\bibfnamefont{G.}~\bibnamefont{Pagliara}},
  \bibinfo{journal}{Astrophys.J} \textbf{\bibinfo{volume}{665}},
  \bibinfo{pages}{1227} (\bibinfo{year}{2007}), \eprint{astro-ph/0512602}.

\bibitem[{\citenamefont{Drago et~al.}(2007{\natexlab{b}})\citenamefont{Drago,
  Lavagno, and Parenti}}]{parenti}
\bibinfo{author}{\bibfnamefont{A.}~\bibnamefont{Drago}},
  \bibinfo{author}{\bibfnamefont{A.}~\bibnamefont{Lavagno}}, \bibnamefont{and}
  \bibinfo{author}{\bibfnamefont{I.}~\bibnamefont{Parenti}},
  \bibinfo{journal}{Astrophys.J} \textbf{\bibinfo{volume}{659}},
  \bibinfo{pages}{1519} (\bibinfo{year}{2007}{\natexlab{b}}).

\bibitem[{\citenamefont{Baldo et~al.}(2003)}]{Baldo:2002ju}
\bibinfo{author}{\bibfnamefont{M.}~\bibnamefont{Baldo}} \bibnamefont{et~al.},
  \bibinfo{journal}{Phys. Lett.} \textbf{\bibinfo{volume}{B562}},
  \bibinfo{pages}{153} (\bibinfo{year}{2003}), \eprint{nucl-th/0212096}.

\bibitem[{\citenamefont{Buballa et~al.}(2004)\citenamefont{Buballa, Neumann,
  Oertel, and Shovkovy}}]{Buballa:2003et}
\bibinfo{author}{\bibfnamefont{M.}~\bibnamefont{Buballa}},
  \bibinfo{author}{\bibfnamefont{F.}~\bibnamefont{Neumann}},
  \bibinfo{author}{\bibfnamefont{M.}~\bibnamefont{Oertel}}, \bibnamefont{and}
  \bibinfo{author}{\bibfnamefont{I.}~\bibnamefont{Shovkovy}},
  \bibinfo{journal}{Phys. Lett.} \textbf{\bibinfo{volume}{B595}},
  \bibinfo{pages}{36} (\bibinfo{year}{2004}), \eprint{nucl-th/0312078}.

\bibitem[{\citenamefont{Klahn et~al.}(2006)}]{Klahn:2006iw}
\bibinfo{author}{\bibfnamefont{T.}~\bibnamefont{Klahn}} \bibnamefont{et~al.}
  (\bibinfo{year}{2006}), \eprint{nucl-th/0609067}.

\bibitem[{\citenamefont{Grunfeld et~al.}(2007)}]{Grunfeld:2007jt}
\bibinfo{author}{\bibfnamefont{A.~G.} \bibnamefont{Grunfeld}}
  \bibnamefont{et~al.} (\bibinfo{year}{2007}), \eprint{arXiv:0705.3787
  [hep-ph]}.

\bibitem[{\citenamefont{Lawley et~al.}(2006)\citenamefont{Lawley, Bentz, and
  Thomas}}]{Lawley:2006ps}
\bibinfo{author}{\bibfnamefont{S.}~\bibnamefont{Lawley}},
  \bibinfo{author}{\bibfnamefont{W.}~\bibnamefont{Bentz}}, \bibnamefont{and}
  \bibinfo{author}{\bibfnamefont{A.~W.} \bibnamefont{Thomas}},
  \bibinfo{journal}{J. Phys.} \textbf{\bibinfo{volume}{G32}},
  \bibinfo{pages}{667} (\bibinfo{year}{2006}), \eprint{nucl-th/0602014}.

\bibitem[{\citenamefont{Baldo et~al.}(2007)\citenamefont{Baldo, Burgio,
  Castorina, Plumari, and Zappala}}]{Baldo:2006bt}
\bibinfo{author}{\bibfnamefont{M.}~\bibnamefont{Baldo}},
  \bibinfo{author}{\bibfnamefont{G.~F.} \bibnamefont{Burgio}},
  \bibinfo{author}{\bibfnamefont{P.}~\bibnamefont{Castorina}},
  \bibinfo{author}{\bibfnamefont{S.}~\bibnamefont{Plumari}}, \bibnamefont{and}
  \bibinfo{author}{\bibfnamefont{D.}~\bibnamefont{Zappala}},
  \bibinfo{journal}{Phys. Rev.} \textbf{\bibinfo{volume}{C75}},
  \bibinfo{pages}{035804} (\bibinfo{year}{2007}), \eprint{hep-ph/0607343}.

\bibitem[{\citenamefont{Alford and Reddy}(2003)}]{Alford:2002rj}
\bibinfo{author}{\bibfnamefont{M.}~\bibnamefont{Alford}} \bibnamefont{and}
  \bibinfo{author}{\bibfnamefont{S.}~\bibnamefont{Reddy}},
  \bibinfo{journal}{Phys. Rev.} \textbf{\bibinfo{volume}{D67}},
  \bibinfo{pages}{074024} (\bibinfo{year}{2003}), \eprint{nucl-th/0211046}.

\bibitem[{\citenamefont{Drago et~al.}(2004)\citenamefont{Drago, Lavagno, and
  Pagliara}}]{Drago:2004vu}
\bibinfo{author}{\bibfnamefont{A.}~\bibnamefont{Drago}},
  \bibinfo{author}{\bibfnamefont{A.}~\bibnamefont{Lavagno}}, \bibnamefont{and}
  \bibinfo{author}{\bibfnamefont{G.}~\bibnamefont{Pagliara}},
  \bibinfo{journal}{Phys. Rev.} \textbf{\bibinfo{volume}{D69}},
  \bibinfo{pages}{057505} (\bibinfo{year}{2004}), \eprint{nucl-th/0401052}.

\bibitem[{\citenamefont{Bombaci et~al.}(2007)\citenamefont{Bombaci, Lugones,
  and Vidana}}]{Bombaci:2006cs}
\bibinfo{author}{\bibfnamefont{I.}~\bibnamefont{Bombaci}},
  \bibinfo{author}{\bibfnamefont{G.}~\bibnamefont{Lugones}}, \bibnamefont{and}
  \bibinfo{author}{\bibfnamefont{I.}~\bibnamefont{Vidana}},
  \bibinfo{journal}{Astron. Astrophys.} \textbf{\bibinfo{volume}{462}},
  \bibinfo{pages}{1017} (\bibinfo{year}{2007}), \eprint{astro-ph/0603644}.

\bibitem[{\citenamefont{Alford and Rajagopal}(2002)}]{Alford:2002kj}
\bibinfo{author}{\bibfnamefont{M.}~\bibnamefont{Alford}} \bibnamefont{and}
  \bibinfo{author}{\bibfnamefont{K.}~\bibnamefont{Rajagopal}},
  \bibinfo{journal}{JHEP} \textbf{\bibinfo{volume}{06}}, \bibinfo{pages}{031}
  (\bibinfo{year}{2002}), \eprint{hep-ph/0204001}.

\bibitem[{\citenamefont{Ruester
  et~al.}(2006{\natexlab{b}})\citenamefont{Ruester, Hempel, and
  Schaffner-Bielich}}]{Ruester:2005fm}
\bibinfo{author}{\bibfnamefont{S.~B.} \bibnamefont{Ruester}},
  \bibinfo{author}{\bibfnamefont{M.}~\bibnamefont{Hempel}}, \bibnamefont{and}
  \bibinfo{author}{\bibfnamefont{J.}~\bibnamefont{Schaffner-Bielich}},
  \bibinfo{journal}{Phys. Rev.} \textbf{\bibinfo{volume}{C73}},
  \bibinfo{pages}{035804} (\bibinfo{year}{2006}{\natexlab{b}}),
  \eprint{astro-ph/0509325}.

\bibitem[{\citenamefont{Baym et~al.}(1971)\citenamefont{Baym, Pethick, and
  Sutherland}}]{Baym:1971pw}
\bibinfo{author}{\bibfnamefont{G.}~\bibnamefont{Baym}},
  \bibinfo{author}{\bibfnamefont{C.}~\bibnamefont{Pethick}}, \bibnamefont{and}
  \bibinfo{author}{\bibfnamefont{P.}~\bibnamefont{Sutherland}},
  \bibinfo{journal}{Astrophys. J.} \textbf{\bibinfo{volume}{170}},
  \bibinfo{pages}{299} (\bibinfo{year}{1971}).

\bibitem[{\citenamefont{Glendenning and Moszkowski}(1991)}]{Glendenning:1991es}
\bibinfo{author}{\bibfnamefont{N.~K.} \bibnamefont{Glendenning}}
  \bibnamefont{and} \bibinfo{author}{\bibfnamefont{S.~A.}
  \bibnamefont{Moszkowski}}, \bibinfo{journal}{Phys. Rev. Lett.}
  \textbf{\bibinfo{volume}{67}}, \bibinfo{pages}{2414} (\bibinfo{year}{1991}).

\bibitem[{\citenamefont{Seidov}(1971)}]{1971AZh....48..443S}
\bibinfo{author}{\bibfnamefont{Z.~F.} \bibnamefont{Seidov}},
  \bibinfo{journal}{Ast. Zh.} \textbf{\bibinfo{volume}{48}},
  \bibinfo{pages}{443} (\bibinfo{year}{1971}).

\bibitem[{\citenamefont{Zdunik et~al.}(1987)\citenamefont{Zdunik, Haensel, and
  Schaeffer}}]{1987A&A...172...95Z}
\bibinfo{author}{\bibfnamefont{J.~L.} \bibnamefont{Zdunik}},
  \bibinfo{author}{\bibfnamefont{P.}~\bibnamefont{Haensel}}, \bibnamefont{and}
  \bibinfo{author}{\bibfnamefont{R.}~\bibnamefont{Schaeffer}},
  \bibinfo{journal}{Astron. Astrophys.} \textbf{\bibinfo{volume}{172}},
  \bibinfo{pages}{95} (\bibinfo{year}{1987}).

\bibitem[{\citenamefont{Lindblom}(1998)}]{Lindblom:1998dp}
\bibinfo{author}{\bibfnamefont{L.}~\bibnamefont{Lindblom}},
  \bibinfo{journal}{Phys. Rev.} \textbf{\bibinfo{volume}{D58}},
  \bibinfo{pages}{024008} (\bibinfo{year}{1998}), \eprint{gr-qc/9802072}.

\bibitem[{\citenamefont{Kaempfer}(1981)}]{kaempfer}
\bibinfo{author}{\bibfnamefont{B.}~\bibnamefont{Kaempfer}},
  \bibinfo{journal}{Physics Letters B} \textbf{\bibinfo{volume}{101}},
  \bibinfo{pages}{366} (\bibinfo{year}{1981}).

\bibitem[{\citenamefont{Hanauske et~al.}(2001)\citenamefont{Hanauske, Satarov,
  Mishustin, Stoecker, and Greiner}}]{Hanauske:2001nc}
\bibinfo{author}{\bibfnamefont{M.}~\bibnamefont{Hanauske}},
  \bibinfo{author}{\bibfnamefont{L.~M.} \bibnamefont{Satarov}},
  \bibinfo{author}{\bibfnamefont{I.~N.} \bibnamefont{Mishustin}},
  \bibinfo{author}{\bibfnamefont{H.}~\bibnamefont{Stoecker}}, \bibnamefont{and}
  \bibinfo{author}{\bibfnamefont{W.}~\bibnamefont{Greiner}},
  \bibinfo{journal}{Phys. Rev.} \textbf{\bibinfo{volume}{D64}},
  \bibinfo{pages}{043005} (\bibinfo{year}{2001}), \eprint{astro-ph/0101267}.

\bibitem[{\citenamefont{Manka and Przybyla}(2002)}]{Manka:2002yv}
\bibinfo{author}{\bibfnamefont{R.}~\bibnamefont{Manka}} \bibnamefont{and}
  \bibinfo{author}{\bibfnamefont{G.}~\bibnamefont{Przybyla}},
  \bibinfo{journal}{New J. Phys.} \textbf{\bibinfo{volume}{4}},
  \bibinfo{pages}{14} (\bibinfo{year}{2002}), \eprint{nucl-th/0201003}.

\bibitem[{\citenamefont{Buballa}(2005)}]{Buballa:2003qv}
\bibinfo{author}{\bibfnamefont{M.}~\bibnamefont{Buballa}},
  \bibinfo{journal}{Phys. Rept.} \textbf{\bibinfo{volume}{407}},
  \bibinfo{pages}{205} (\bibinfo{year}{2005}), \eprint{hep-ph/0402234}.

\bibitem[{\citenamefont{Buballa and Oertel}(1999)}]{Buballa:1998pr}
\bibinfo{author}{\bibfnamefont{M.}~\bibnamefont{Buballa}} \bibnamefont{and}
  \bibinfo{author}{\bibfnamefont{M.}~\bibnamefont{Oertel}},
  \bibinfo{journal}{Phys. Lett.} \textbf{\bibinfo{volume}{B457}},
  \bibinfo{pages}{261} (\bibinfo{year}{1999}), \eprint{hep-ph/9810529}.

\bibitem[{\citenamefont{Schertler et~al.}(1999)\citenamefont{Schertler,
  Leupold, and Schaffner-Bielich}}]{Schertler:1999xn}
\bibinfo{author}{\bibfnamefont{K.}~\bibnamefont{Schertler}},
  \bibinfo{author}{\bibfnamefont{S.}~\bibnamefont{Leupold}}, \bibnamefont{and}
  \bibinfo{author}{\bibfnamefont{J.}~\bibnamefont{Schaffner-Bielich}},
  \bibinfo{journal}{Phys. Rev.} \textbf{\bibinfo{volume}{C60}},
  \bibinfo{pages}{025801} (\bibinfo{year}{1999}), \eprint{astro-ph/9901152}.

\bibitem[{\citenamefont{Hatta and Fukushima}(2004)}]{Hatta:2003ga}
\bibinfo{author}{\bibfnamefont{Y.}~\bibnamefont{Hatta}} \bibnamefont{and}
  \bibinfo{author}{\bibfnamefont{K.}~\bibnamefont{Fukushima}},
  \bibinfo{journal}{Phys. Rev.} \textbf{\bibinfo{volume}{D69}},
  \bibinfo{pages}{097502} (\bibinfo{year}{2004}), \eprint{hep-ph/0307068}.

\bibitem[{\citenamefont{Bender et~al.}(1998)\citenamefont{Bender, Poulis,
  Roberts, Schmidt, and Thomas}}]{Bender:1997jf}
\bibinfo{author}{\bibfnamefont{A.}~\bibnamefont{Bender}},
  \bibinfo{author}{\bibfnamefont{G.~I.} \bibnamefont{Poulis}},
  \bibinfo{author}{\bibfnamefont{C.~D.} \bibnamefont{Roberts}},
  \bibinfo{author}{\bibfnamefont{S.~M.} \bibnamefont{Schmidt}},
  \bibnamefont{and} \bibinfo{author}{\bibfnamefont{A.~W.}
  \bibnamefont{Thomas}}, \bibinfo{journal}{Phys. Lett.}
  \textbf{\bibinfo{volume}{B431}}, \bibinfo{pages}{263} (\bibinfo{year}{1998}),
  \eprint{nucl-th/9710069}.

\bibitem[{\citenamefont{Fraga et~al.}(2001)\citenamefont{Fraga, Pisarski, and
  Schaffner-Bielich}}]{Fraga:2001id}
\bibinfo{author}{\bibfnamefont{E.~S.} \bibnamefont{Fraga}},
  \bibinfo{author}{\bibfnamefont{R.~D.} \bibnamefont{Pisarski}},
  \bibnamefont{and}
  \bibinfo{author}{\bibfnamefont{J.}~\bibnamefont{Schaffner-Bielich}},
  \bibinfo{journal}{Phys. Rev.} \textbf{\bibinfo{volume}{D63}},
  \bibinfo{pages}{121702} (\bibinfo{year}{2001}), \eprint{hep-ph/0101143}.

\bibitem[{\citenamefont{McLerran and Pisarski}(2007)}]{McLerran:2007qj}
\bibinfo{author}{\bibfnamefont{L.}~\bibnamefont{McLerran}} \bibnamefont{and}
  \bibinfo{author}{\bibfnamefont{R.~D.} \bibnamefont{Pisarski}}
  (\bibinfo{year}{2007}), \eprint{arXiv:0706.2191 [hep-ph]}.

\bibitem[{\citenamefont{Drago et~al.}(2007{\natexlab{c}})\citenamefont{Drago,
  Pagliara, and Schaffner-Bielich}}]{Drago:2007zk}
\bibinfo{author}{\bibfnamefont{A.}~\bibnamefont{Drago}},
  \bibinfo{author}{\bibfnamefont{G.}~\bibnamefont{Pagliara}}, \bibnamefont{and}
  \bibinfo{author}{\bibfnamefont{J.}~\bibnamefont{Schaffner-Bielich}}
  (\bibinfo{year}{2007}{\natexlab{c}}), \eprint{arXiv:0705.4418 [astro-ph]}.

\bibitem[{\citenamefont{Neumann et~al.}(2003)\citenamefont{Neumann, Buballa,
  and Oertel}}]{Neumann:2002jm}
\bibinfo{author}{\bibfnamefont{F.}~\bibnamefont{Neumann}},
  \bibinfo{author}{\bibfnamefont{M.}~\bibnamefont{Buballa}}, \bibnamefont{and}
  \bibinfo{author}{\bibfnamefont{M.}~\bibnamefont{Oertel}},
  \bibinfo{journal}{Nucl. Phys.} \textbf{\bibinfo{volume}{A714}},
  \bibinfo{pages}{481} (\bibinfo{year}{2003}), \eprint{hep-ph/0210078}.

\bibitem[{\citenamefont{Kitazawa et~al.}(2007)\citenamefont{Kitazawa, Rischke,
  and Shovkovy}}]{Kitazawa:2007zs}
\bibinfo{author}{\bibfnamefont{M.}~\bibnamefont{Kitazawa}},
  \bibinfo{author}{\bibfnamefont{D.~H.} \bibnamefont{Rischke}},
  \bibnamefont{and} \bibinfo{author}{\bibfnamefont{I.~A.}
  \bibnamefont{Shovkovy}} (\bibinfo{year}{2007}), \eprint{arXiv:0709.2235
  [hep-ph]}.

\bibitem[{\citenamefont{Nickel et~al.}(2006)\citenamefont{Nickel, Alkofer, and
  Wambach}}]{Nickel:2006kc}
\bibinfo{author}{\bibfnamefont{D.}~\bibnamefont{Nickel}},
  \bibinfo{author}{\bibfnamefont{R.}~\bibnamefont{Alkofer}}, \bibnamefont{and}
  \bibinfo{author}{\bibfnamefont{J.}~\bibnamefont{Wambach}},
  \bibinfo{journal}{Phys. Rev.} \textbf{\bibinfo{volume}{D74}},
  \bibinfo{pages}{114015} (\bibinfo{year}{2006}), \eprint{hep-ph/0609198}.

\bibitem[{\citenamefont{Ippolito
  et~al.}(2007{\natexlab{b}})\citenamefont{Ippolito, Ruggieri, Rischke,
  Sedrakian, and Weber}}]{nicola}
\bibinfo{author}{\bibfnamefont{N.}~\bibnamefont{Ippolito}},
  \bibinfo{author}{\bibfnamefont{M.}~\bibnamefont{Ruggieri}},
  \bibinfo{author}{\bibfnamefont{D.~H.} \bibnamefont{Rischke}},
  \bibinfo{author}{\bibfnamefont{A.}~\bibnamefont{Sedrakian}},
  \bibnamefont{and} \bibinfo{author}{\bibfnamefont{F.}~\bibnamefont{Weber}}
  (\bibinfo{year}{2007}{\natexlab{b}}), \eprint{arXiv:0710.3874 [astro-ph]}.

\bibitem[{\citenamefont{Schertler et~al.}(2000)\citenamefont{Schertler,
  Greiner, Schaffner-Bielich, and Thoma}}]{Schertler:2000xq}
\bibinfo{author}{\bibfnamefont{K.}~\bibnamefont{Schertler}},
  \bibinfo{author}{\bibfnamefont{C.}~\bibnamefont{Greiner}},
  \bibinfo{author}{\bibfnamefont{J.}~\bibnamefont{Schaffner-Bielich}},
  \bibnamefont{and} \bibinfo{author}{\bibfnamefont{M.~H.} \bibnamefont{Thoma}},
  \bibinfo{journal}{Nucl. Phys.} \textbf{\bibinfo{volume}{A677}},
  \bibinfo{pages}{463} (\bibinfo{year}{2000}), \eprint{astro-ph/0001467}.

\bibitem[{\citenamefont{Glendenning and Kettner}(2000)}]{Glendenning:1998ag}
\bibinfo{author}{\bibfnamefont{N.~K.} \bibnamefont{Glendenning}}
  \bibnamefont{and} \bibinfo{author}{\bibfnamefont{C.}~\bibnamefont{Kettner}},
  \bibinfo{journal}{Astron. Astrophys.} \textbf{\bibinfo{volume}{353}},
  \bibinfo{pages}{L9} (\bibinfo{year}{2000}), \eprint{astro-ph/9807155}.

\end{thebibliography}
\bibliographystyle{apsrev}
\end{document}